# Disorder-assisted Robustness of Ultrafast Cooling in High Doped CVD-Graphene

Tingyuan Jia[1,2,3], Wenjie Zhang[4], Zijun Zhan[1,2], Zeyu Zhang[1,2,3]\*, Guohong Ma[4], Juan Du[1,2,3]\* and Yuxin Leng[1,2,3]\*

[1] *State Key Laboratory of High Field Laser Physics and CAS Center for Excellence in Ultra-intense Laser Science, Shanghai Institute of Optics and Fine Mechanics (SIOM), Chinese Academy of Sciences (CAS), Shanghai 201800, China*

[2] *Hangzhou Institute for Advanced Study, University of Chinese Academy of Sciences, Hangzhou，310024, China*

[3] *Center of Materials Science and Optoelectronics Engineering, University of Chinese Academy of Sciences, Beijing 100049, China*

[4] *Department of Physics, Shanghai University, 99 Shangda Road, Shanghai 200444, China*

*†These authors contribute equally to this work.*
*\*Corresponding author: zhangzeyu@siom.ac.cn, dujuan@siom.ac.cn, lengyuxin@siom.ac.cn*

**Abstract:**

Dirac Fermion, which is the low energy collective excitation near the Dirac cone in monolayer graphene, have gained great attention by low energy Terahertz probe. In the case of undoped graphene, it has been generally understood that the ultrafast terahertz thermal relaxation is mostly driven by the electron-phonon coupling (EOP), which can be prolonged to tens and hundreds of picoseconds. However, for the high doped graphene, which manifests the negative photoinduced terahertz conductivity, there is still no consensus on the dominant aspects of the cooling process on a time scale of a few picoseconds. Here, the competition between the disorders assisted defect scattering and the electron-phonon coupling process in the cooling process of the graphene terahertz dynamics is systematically studied and disentangled. We verify experimentally that the ultrafast disorder assisted lattice-phonon interaction, rather than the electron-phonon coupling process, would play the key role in the ultrafast thermal relaxation of the terahertz dynamics. Furthermore, the cooling process features robustness which is independent on the pump wavelength and external temperature. Our finding is expected to propose a considerable possible cooling channel in CVD-graphene and to increase the hot electron extracting efficiency for the design of graphene-based photoconversion devices.

The energy below absorber band gap (Eg) is not captured and any excess energy exceeding Eg is lost irreversibly to lattice as heat, leading to Shockley-Queisser limit for photovoltaic cells.[1] One way to overcome this limit and achieve far higher efficiency is absorbing a broadband spectrum and extracting the hot electrons before the electron cooling process.[2-4] However, the cooling mechanism of the hot Dirac Fermion detected by Terahertz wave in graphene are still controversial, but will significantly affect its ultrafast photo-induced performance.[5-8] According to the two temperature model, the thermalization and relaxation dynamics in graphene has been well interpreted in term of electron-electron and electron-phonon coupling, with the appearance of the electron-optical phonon coupling, the electron-acoustic phonon supercollision process.[9-15] However, for the highly doped CVD grown monolayer graphene, the photoinduced negative terahertz conductivity cooling mechanism have been much more complicated and lack of consensus.[8,10,16-19] Currently, three different models have been proposed to explain the ultrafast terahertz relaxation dynamics in CVD-grown graphene. In 2012, the electron-acoustic phonon supercollision model has been proposed, which explains why the thermal relaxation in CVD-grown graphene is much faster than the dynamics in bulk semi-conductor.[10,20] Later, the thermodynamic model, in which the intraband thermalization is induced both by the electron-electron coupling and the electron-phonon coupling process, could explain the slowing down behavior of the terahertz conductivity relaxation in graphene successfully.[17,18,21] However, the disorder-assisted acoustic phonon scattering and out-of-plane heat transfer to the environment has been proved to be inefficient, Minev et. al shows that the relaxation dynamics could be dominated by the electron-optical phonon coupling without the electron-acoustic phonon supercollision in high doping graphene and Eva Pogna et. al claimed that the optical-to-acoustic phonon coupling dominate the hot carrier cooling dynamics in high-quality graphene.[19,22]

Even though controversial views have arisen from the three models mentioned above, it could be concluded that the electron phonon coupling tuned by the Fermi level and the degree of disorder or defect, could act as possible key factors determining the ultrafast conductivity relaxation in CVD-graphene. It is notably that the Fermi level and the degree of disorder of the sample could be affected by the substrate,[6,7] and the thermalization rate and the electron-phonon coupling in CVD-graphene could be modulated by the pump fluence and the pump photon energy.[16] In addition, the population of the acoustic and optical phonons, which is crucial for the supercollison process, could be modulated by the temperature of the graphene film.[15,16] Therefore, to clarify the controversial contribution to the ultrafast thermal relaxation of these factors, the most straightforward first step seems to carry on a comprehensive tunable experiment to investigate influence of the substrates (disorders), pump wavelength, duration, fluence and the external temperature of graphene, which is vital to provide a fundamental understanding of the ultrafast THz conductivity cooling dynamics in CVD graphene.

Here, we disentangle the initial ultrafast terahertz conductivity dynamics in monolayer CVD-graphene by constructing quantitatively degree of disorder, carrier density, phonon density and initial hot electron energy using different substrates, pump influences and pump-probe wavelengths at variable temperatures, respectively. We reveal that after the lattice temperature thermalization by the electron-optical phonon scattering process in the first ~100 fs, the subsequent relaxation process in around 1~3 picoseconds are independent on the population of the acoustic phonon as the temperature changes. In addition, the saturation relaxation time does not change with the pump photon energy and temperature, but strongly dominated by the degree of the disorder in monolayer CVD-graphene. Therefore, the disorder assisted cooling process is supposed to be the most possible ultrafast relaxation pathway in CVD grown monolayer graphene. Our results reveled the crucial mechanism of the ultrafast electron cooling process for the hot electron hole harvesting in graphene. This finding represented a promising step toward a new application of active photonic and phononic devices based on CVD-grown graphene.

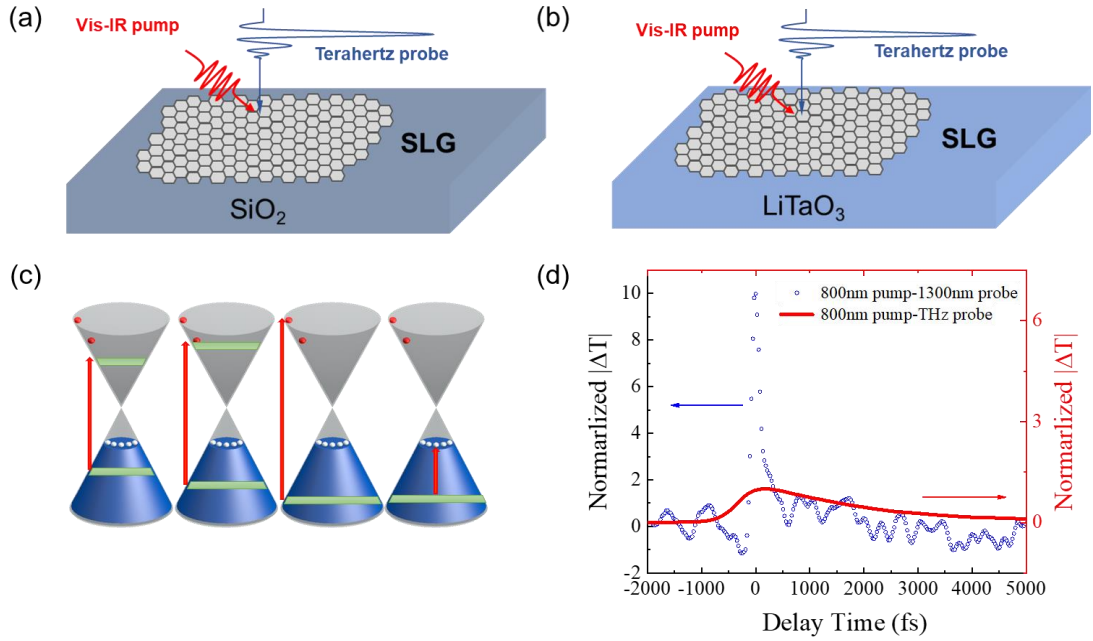

Figure 1 (a) The schematic diagram of graphene thin film on SiO$_2$ substrate; (b) The schematic diagram of graphene thin film on LiTaO$_3$ substrate. (c) The transformation of the Fermi surface during the ultrafast thermalization and cooling process.; (d) Normalized |ΔT| as a function of delay time for graphene/SiO$_2$ by infrared and THz probe, respectively.

As shown in Fig.1, CVD-grown graphene is transferred to SiO$_2$ and LiTaO$_3$ substrates, respectively. The transient transmittance and the photo-induced terahertz conductivity dynamics of graphene are presented by using 800nm (1.55 eV) laser pump, 1300nm (0.95 eV) and terahertz probe, respectively. As shown in Fig. 1 (d), despite the totally different probe wavelength in these two experiments, the negative photo-induced conductivity relaxation dynamics is well coinciding with the slow relaxation dynamic infrared probe trace after electron-phonon relaxation. In the case of 35 fs infrared pulse probe, after the vividly rising process of the transmittance signal, there is a fast relaxation process within 0.2 ps and a relatively slow process of 1~3 ps (Figure 2 (a)&(b)). The similar ultrafast relaxation process in the order of 200 femtoseconds is also observed for hot carriers in the free-standing graphene using time-resolved angle-resolved photoelectron spectroscopy.[13] However, as shown in Fig. 1 (d) and S1, there is only one exponential decay process in terahertz conductivity relaxation, because the electron thermalization and the subsequent electron-phonon thermalization process are swallowed by the THz pulse with duration about 1 ps longer than them. Therefore, optical excitation and probe with shorter pulse duration combined with tunable pump-THz probe spectroscopy is highly desired to confirm the one-exponential conductivity relaxation process in CVD-graphene is whether consisted with the electron-optical phonon coupling.

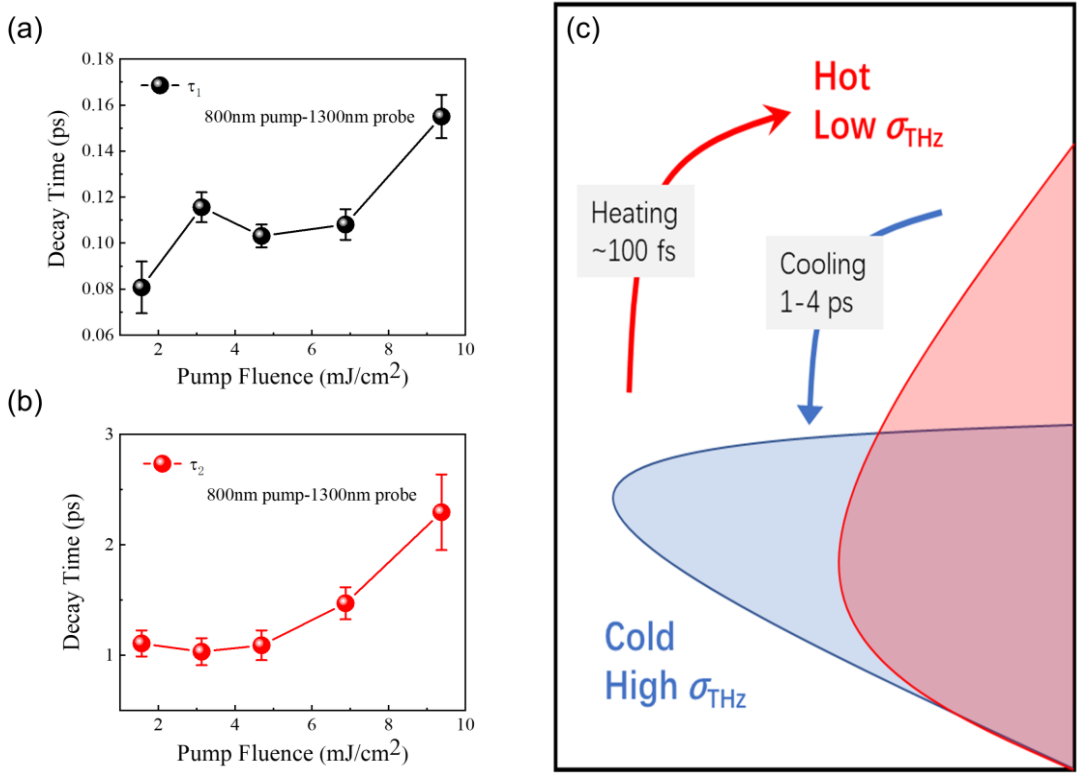

Figure 2 (a) and (b) The extracted values for the relaxation process of graphene/SiO$_2$ as a function of pump fluence. (c) Hot-carrier relaxation and cooling dynamics in graphene/SiO$_2$.

The photo-induced terahertz conductivity in CVD-grown graphene is negative and could be regarded as the reduction of the intraband scattering rate of the electron near the Fermi surface.[19,23] As shown in Fig. 1 (d), the negative terahertz conductivity cooling should correspond to the slow process of the 0.95eV probe trace. As shown in Fig. 2, two different processes are extracted from the 0.95eV probe dynamics by the two-exponential fitting. The typical decay of the transient dynamics can be described by the phenomenologically three temperature model.[13,15,21,24,25]

$$\frac{\partial T_l}{\partial t} = \frac{C_e}{C_l}(9.62 \frac{\lambda_2 g k_B^3}{\hbar k_F L} \frac{T_e^3 - T_l^3}{C_e} + \pi \lambda_2 \hbar g[\mu(Te)] k_F^2 v_s^2 k_B \frac{T_e - T_l}{C_e}), \qquad (1)$$

$$\frac{\partial T}{\partial t} = +\frac{s(t)}{\beta} - \frac{\pi \lambda_1 g[\mu(Te)]\Omega^3}{\hbar^-} \frac{n_e - n_p}{Ce} - 9.62 \frac{\lambda_y g[u((Te))] k_B^3}{\hbar k_F l} \frac{T_e^3 - T_l^3}{Ce} - \pi \lambda_2 \hbar g[\mu(Te)] k_f^2 u_s^2 k_B \frac{Te - Tl}{Ce}, \qquad (2)$$

$$\frac{\partial T_P}{\partial t} = \frac{C_e}{C_P} \frac{\pi \lambda_1 g[u(Te)]\Omega_E^3}{\hbar} \frac{n_e - n}{C_e} \qquad (3)$$

Where $T_e$ means electron temperature, $T_p$ means phonon temperature and $T_l$ means lattice temperature. From the formula, the lattice relaxation process is strongly dependent on the density of the electron state $g[u((Te))]$, the scaling parameter $\beta$ and the electron-phonon coupling constants, $\lambda_1$ and $\lambda_2$. Here, $g$ consists of the Fermi energy, and the electron-acoustic phonon coupling constant $\lambda_2$ is consisted with the disorder of the graphene. According to the formula (2), the phonon cooling rate is in perpendicular to the temperature of the electron $T_e$. As shown in Fig.2 (b), when the pump fluence increases, the $T_e$ increases vividly, resulting in the increase of the electron-lattice thermalization time, which is consistent with the formula. While for the phonon relaxation as described in formula (3), the relaxation time should be strongly dependent on the optical phonon energy $\Omega_E$. It has been reported that $\Omega_E$ is 200meV for graphene and the optical phonon density is independent on the electron density of the state.[26,27] Therefore, it is crucial to measure the ultrafast terahertz dynamics by tunable pump photon energy, especially those near $2\Omega_E$=0.4eV, to verify the role of electron-optical phonon coupling in the relaxation process.

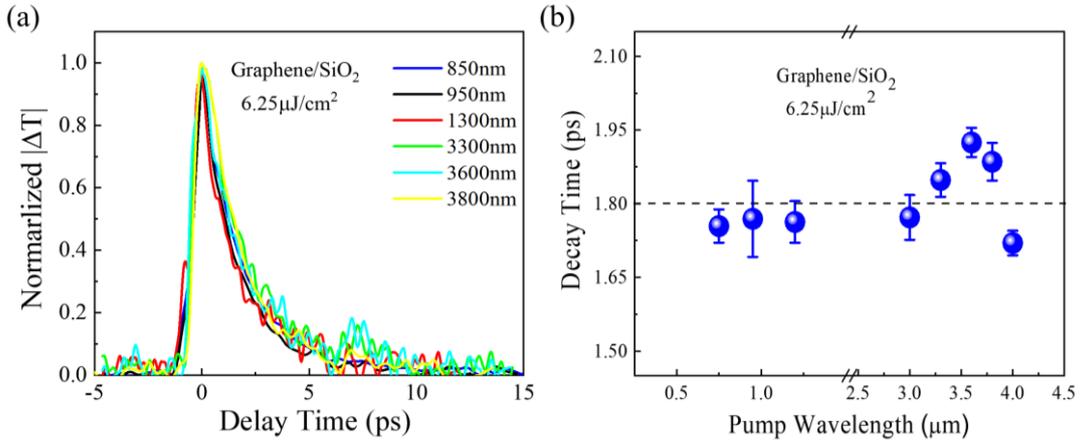

Figure 3 (a) The pump fluence dependence of the relaxation process of graphene/SiO$_2$ with different photon energy. (b) Normalized |ΔT| as a function of delay time for graphene/SiO$_2$ by different photon energy at a fixed pump fluence of 6.25μJ/cm². (c) The ultrafast cooling time constant of graphene/SiO$_2$ with different photon energy at a fixed pump fluence of 6.25μJ/cm².

Figure 3 (a) shows the terahertz conductivity relaxation time dependence on the pump fluence and pump photon energy. The decay time for each pump fluence is slightly differentiated due to the different thermal efficiency for various photon energy pumping.[14,16] However, when the pump fluence reached saturation (after 300uJ/cm²) for all the pump photon energy, the decay time comes to the same time scale for about 2.8 picoseconds, which shows relaxation robustness for the pump photon energy. The ultrafast relaxation process of the negative terahertz dynamics is considered to be related to the electron-optical phonon coupling in the microscopic understanding.[19] If so, different pump photon energy would lead to thermalization of the graphene lattice with different efficiency by means of the so called carrier multiplication as shown in Fig. 1 (c).[14] In other word, the decay time should be linearly proportional to the pump photon energy. However, as shown in Fig. 3 (a) and Fig. S1, the peak before saturation is not linearly dependent on pump photon energy, and the decay time almost remains constant after saturation. Consequently, the ultrafast relaxation process of the negative terahertz dynamics is difficult to be explained by the carrier multiplication via the electron optical phonon coupling. To verify the ultrafast relaxation process of the negative terahertz dynamics further, pump photon energy in the range of 0.3eV~0.4eV (3~4 μm) is employed to eliminate the influence of the optical phonon. Since the pump photon energy is smaller than the 2$\Omega_E$, the instant hot electron does not have enough energy to interact with the optical phonon in graphene.[26] As shown in Fig.3 (b), when the pump photon energy comes to 4μm (0.31 eV), the relaxation process also falls on the same time scale compared with other pump photon energies in about 1.8 picoseconds in an unsaturable condition at a pump fluence of 6.25uJ/cm$^2$.[23] This feature reconfirmed the electron-optical phonon coupling is relatively inefficient in the ultrafast terahertz conductivity relaxation of high doped graphene. Therefore, the role of the electron-phonon coupling process (both for the optical and acoustic phonon) in the conductivity cooling process should be reconsidered.

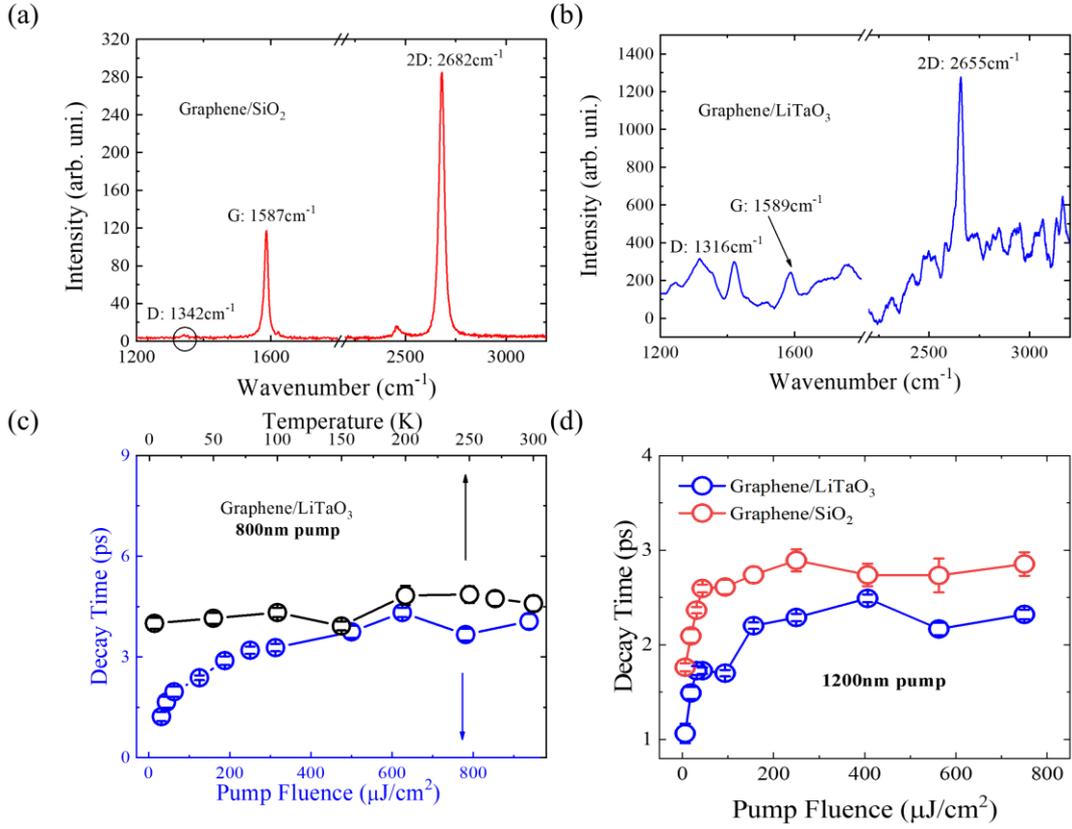

Figure 4 (a) and (b) Raman spectroscopy of CVD-grown graphene on the fused silica and LiTaO$_3$ substrates, respectively. The relative intensities between the 2D and G peaks identify the monolayer of the graphene. (c) The extracted values for the relaxation process of graphene/LiTaO$_3$ as a function of pump fluence (blue) and temperature (black). (d) The pump fluence dependence of the relaxation process of graphene on SiO$_2$ (red) and LiTaO$_3$ (blue) substrates, respectively.

Different from optical phonon with large energy about 200meV, the acoustic phonon in graphene has a much less average energy of 4meV.[10] The dependence on the electron-optical phonon coupling of the cooling process takes the electron-acoustic phonon coupling in to account. However, because of the small phonon energy (~4meV) compared with the hot electron (0.155eV-0.827eV) if the electron acoustic phonon coupling dominated the following relaxation dynamics, the super-collision model must be taken into concern.[10,15] It is worth knowing that, both for the Debye model and the Bloch–Grüneisen (BG) model in graphene, the acoustic phonon population is strongly dependent on the temperature. Since both the Debye temperature and the BG temperature in graphene are reported to be higher than 3000K,[23] the phonon population can be regarded to be directly proportional to the lattice temperature in our experiment.[21,28] As shown in Fig. 4 (c), as temperature change from 300-4K, after increasing the pump fluence above $800 \mu J/cm^2$, the decay lifetimes remains nearly the same as the external temperature changes. Meanwhile, as shown in Fig. S3, the ultrafast photoinduced conductivity of graphene increased with temperature decreased. The temperature robustness feature suggests that the small contribution of the electron-acoustic phonon coupling in the ultrafast terahertz conductivity relaxation. Similar results have also been observed in some previous literature, which makes the authors exclude the electron-acoustic phonon super collision process in high doped graphene.[19,29]

Therefore, neither the electron-optical phonon coupling, nor the electron-acoustic phonon super-collision model is appropriate to explain the ultrafast terahertz conductivity cooling in high doped graphene. Then we attempt to introduce the influence of the disorder and the Fermi level on the ultrafast thermal relaxation by the aid of different substrate of the graphene.[10] As shown in Fig. 4, in contrast to the independence on the electron-phonon coupling process, the ultrafast thermalization and cooling process in CVD graphene at ultrafast time scale have strongly reliance upon the substrate. In detail, as shown in Fig. 4 (d), as the pump fluence change from $6.25\text{-}750 \mu J/cm^2$ the decay lifetimes gradually increase, and then reach saturation. We notice that the saturated cooling lifetime in graphene/LiTaO$_3$ is shorter than that in graphene/SiO$_2$. Fig 4 (a) and (b) show the Raman spectrum of our CVD-grown graphene on the fused silica substrate and LiTaO$_3$ substrate in air. The mean Fermi energy could be determined to be 167 meV for graphene/SiO$_2$ and 214 meV for graphene/LiTaO$_3$, respectively, using the equation of $|E_F| = (v_G - 1580)/42 \text{cm}^{-1}\text{eV}^{-1}$,[21] where $v_g$ is the vibrational frequency of G mode (1587cm$^{-1}$ for graphene/SiO$_2$ and 1589cm$^{-1}$ for graphene/LiTaO$_3$). It has been reported that when the Fermi energy in CVD-graphene increases, the cooling process will be prolonged due to the multiple electron-phonon coupling relaxation time increases.[18,19,25] While as shown in Fig. 4 (d), the cooling process lifetime in the low Fermi energy sample is longer than that in a relatively high Fermi energy sample, which reconfirm the electron-phonon coupling is not dominant the conductivity cooling process.

As we notice that the intensity ratio of the D model (resulted from defect assisted traverses optical phonon) to the G model (resulted from the monolayer structure of graphene) in graphene/LiTaO$_3$ is much larger than the ratio in graphene/SiO$_2$, as shown in Fig. 4 (a)&(b), which indicates the phonon assisted defect scattering process in graphene/LiTaO$_3$ is more intense than that in graphene/SiO$_2$.[30] In detail, the average density of intrinsic dopant carriers could be calculated to be $1.7 \times 10^{12}\text{cm}^{-2}$ and 2.8×10$^{12}$ cm$^{-2}$ for CVD-grown graphene on fused SiO$_2$ and on LiTaO$_3$, respectively, by $N_i = (E_F)^2/(\hbar^2 V_F^2 \pi)$, where $V_F = 1.1 \times 10^6$ m/s. Therefore, it is reasonable to ascribe the substrate dependence of the ultrafast thermalization and cooling process to the disorder scattering process, which is expected to be the mainly path of the relaxation process and is probably responsible for the phonon recycling and the Zener–Klein tunneling process presented in recent publications.[6,31] In other words, a higher efficient cooling is expected to occur in a higher disordered high doped graphene. However, in fact, it is the lattice temperature that affects the ultrafast terahertz conductivity, so the relaxation of terahertz conductivity can be regarded as the relaxation of lattice temperature. This process can be regarded as the process of lattice vibration diffusion from local to outward. In other words, it is the process of optical phonon-acoustic phonon interaction contribution that could be understanded why this process does not change with temperature. (assuming that both phonons conform to the temperature dependence of Boltzmann distribution). Our finding indicates a reconsideration of the hot electron dynamics applications in areas such as thermal and phonon transistor, terahertz high-harmonic generation devices.

In summary, to disentangle the controversial issue of the high doped graphene terahertz relaxation dynamics and build an intuitive physical picture of the microscopic relaxation model. Through generating the hot electron initial energy less than the minimum optical phonon energy, we find the electron-optical phonon coupling process plays negligible role in the terahertz conductivity relaxation process. In addition, the temperature independence of the cooling process in Graphene/LiTaO$_3$ shows the electron-acoustic phonon supercollision process is not dominated in the cooling process yet. Finally, we find the disorder assist lattice-phonon interaction responsible for the cooling process of the terahertz dynamics. The cooling mechanism is useful for the energy transfer, hot electron extraction and the phonon recycling of graphene heterojunctions, which is the key mechanism for the photo-thermal-electric conversion in the graphene based 2D heterostructure devices. Our results will offer new insights into the application involved with the ultrafast nonequilibrium carriers cooling pathway and the improvement of CVD graphene-based nano-optoelectronic devices.

**Methods and Experimental setup:**

**Ultrafast Spectroscopy:** We performed the VIPTP and OPIP drove by a 1 kHz Ti: Sapphire

regenerative amplifier with 800 nm central wavelength and 35 fs pulse duration. The beam size of the terahertz beam is 0.16 cm$^2$. So, the carrier density could be calculated by the proportion of the absorption photon amount to the irradiation area. The THz pulses are generated by optical rectification and detected by electro-optic sampling in a pair of 1mm thick, (110)-oriented ZnTe crystals. To obtain the terahertz conductivity of the sample, we measure the pump-induced changes of THz peak signals normalized to the THz transmission without photoexcitation ($\Delta T/T_0$) for the CVD graphene samples from different substrates as a function of the pump-probe delay $\Delta t$. As the sample films are thin, $\Delta T/T_0$ is related to the photo-induced sheet photoconductivity, i.e., $\Delta\sigma = \frac{1+n}{Z_0}\left(\frac{1}{1+\Delta T/T_0} - 1\right)$, where n=1.95 is the refractive index of the fused silica substrate, and $Z_0$=377 is the free space impedance.

**Exponential fitting of the carrier dynamics:** The temporal dynamics of the THz conductivity is fitted by:

$$-\Delta\sigma(\Delta t) = A \cdot e^{-\frac{t}{\tau_{decay}}} + C$$

, where $\tau_{decay}$ is the relaxation dynamics of the lattice temperature. C represents the longitude thermal diffusion process which could be regarded as a plateau in our condition.

**Acknowledgement**

This work was supported by National Natural Science Foundation of China (Nos. 92050203, 61875211, 61905264, 61925507, 61521093, 61520106012, 61674023, 62005296), National Key R&D Program of China 2017YFE0123700, The Strategic Priority Research Program of the Chinese Academy of Sciences (No. XDB16030400), the Program of Shanghai Academic/Technology Research Leader (No. 18XD1404200), 2017SHZDZX02, Shanghai Municipal Science and Technology Major Project; Z. J. acknowledge the "Chen Guang" project (16CG45) supported by Shanghai Municipal Education Commission and Shanghai Education Development Foundation.

**Data availability**

The data that support the findings of this study are available from the corresponding author upon reasonable request.

# Disorder-assisted Robustness of Ultrafast Cooling in CVD-Graphene

Tingyuan Jia[1,2,3], Wenjie Zhang[4], Zijun Zhan[1,2], Zeyu Zhang[1,2,3]*, Guohong Ma[4], Juan Du[1,2,3]* and Yuxin Leng[1,2,3]*

[1] *State Key Laboratory of High Field Laser Physics and CAS Center for Excellence in Ultra-intense Laser Science, Shanghai Institute of Optics and Fine Mechanics (SIOM), Chinese Academy of Sciences (CAS), Shanghai 201800, China*

[2] *Hangzhou Institute for Advanced Study, University of Chinese Academy of Sciences, Hangzhou，310024, China*

[3] *Center of Materials Science and Optoelectronics Engineering, University of Chinese Academy of Sciences, Beijing 100049, China*

[4] *Department of Physics, Shanghai University, Shanghai 200444. China*

*†These authors contribute equally to this work.*

*\*Corresponding author: zhangzeyu@siom.ac.cn, dujuan@siom.ac.cn, lengyuxin@siom.ac.cn*

## Supplementary Information

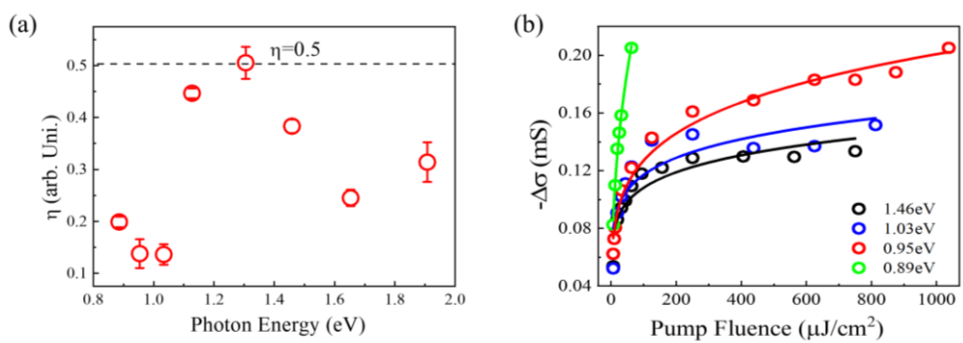

Figure S1. (a) The pump fluence dependence of the conductivity peak with different pump photon energy; (b) The photon energy dependence of the thermal efficiency of the femtosecond laser.

Here, by analyzing the peak value of photo induced conductivity vary with wavelength and the pump fluence, the heating and relaxation process of Dirac fermions in graphene is studied. Figure S1(a) shows the pump fluence dependent peak terahertz photoconductivity of graphene/$SiO_2$ with a power-exponential function fitting. The heating efficiency η can be extracted from the fitting function $\sigma(peak) = AF^\eta$ which is shown in Fig. 3 (b), $\sigma(peak)$ means the peak value of the photo-induced THz conductivity at different pump wavelength (photo energy) and F means the pump fluence.

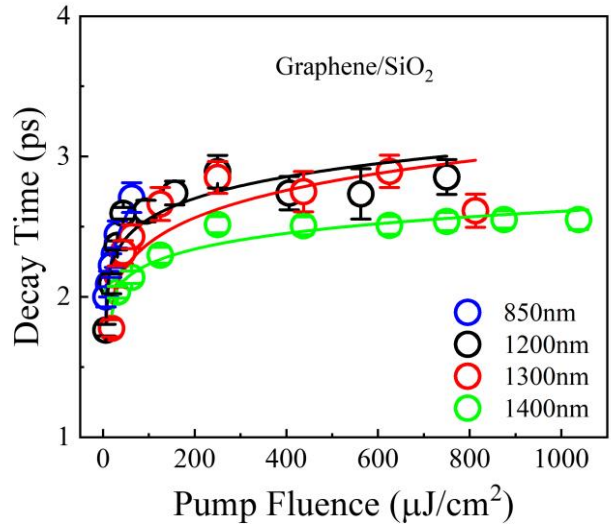

Figure S2. The pump fluence dependence of the relaxation process of graphene/SiO$_2$ with different photon energy.

Figure. S2 shows the terahertz conductivity relaxation time dependence on the pump fluence and pump photon energy. The decay time for each pump fluence is slightly differentiated due to the different thermal efficiency for various photon energy pumping. However, when the pump fluence reached saturation (after 300uJ/cm$^2$) for all the pump photon energy, the decay time comes to the same time scale for about 2.8 picoseconds, which shows relaxation robustness for the pump photon energy.

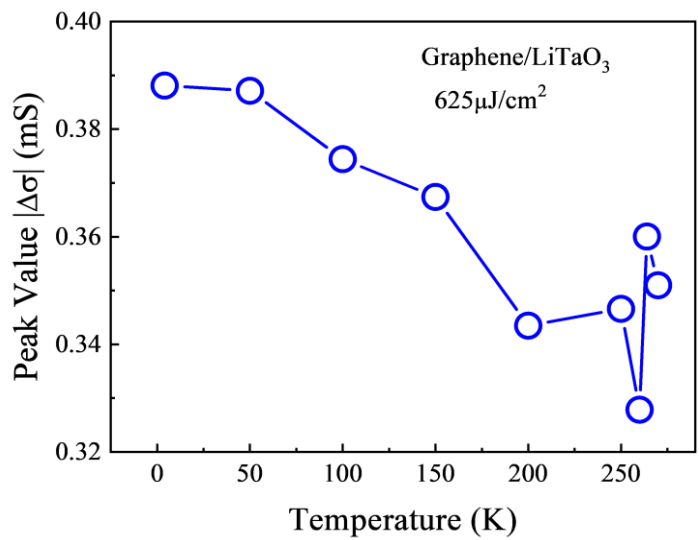

Figure S3. The conductivity peak value |Δσ| of graphene/LiTaO$_3$ as a function of temperature with pump fluence of

As shown in Fig. S3, as temperature varies from 4K to 275K, with the pump fluence at $625 \mu J/cm^2$, the conductivity peak value |Δσ| decreases gradually as the external temperature changes in the range of less than 20%. This feature suggests that the increasing transient conductivity doesn't efficiently affected the relaxation dynamics assuming the higher |Δσ| the slower τ.